\documentclass[twocolumn,preprintnumbers,showpacs,amsmath,amssymb]{revtex4}
\usepackage{graphicx}

\begin{document}
\title{Quantum key distribution based on orthogonal states allows secure
quantum bit commitment}
\author{Guang Ping He}
\email{hegp@mail.sysu.edu.cn}
\affiliation{School of Physics and
Engineering, Sun Yat-sen University, Guangzhou 510275, China}

\begin{abstract}
For more than a decade, it was believed that unconditionally secure
quantum bit commitment (QBC) is impossible. But basing on a
previously proposed quantum key distribution scheme using orthogonal
states, here we build a QBC protocol in which the density matrices
of the quantum states encoding the commitment do not satisfy a
crucial condition on which the no-go proofs of QBC are based. Thus
the no-go proofs could be evaded. Our protocol is fault-tolerant and
very feasible with currently available technology. It reopens the
venue for other \textquotedblleft post-cold-war\textquotedblright\
multi-party cryptographic protocols, e.g., quantum bit string
commitment and quantum strong coin tossing with an arbitrarily small
bias. This result also has a strong influence on the
Clifton-Bub-Halvorson theorem which suggests that quantum theory
could be characterized in terms of information-theoretic
constraints.
\end{abstract}

\maketitle

\newpage

\section{Introduction}

Quantum bit commitment (QBC) is an essential primitive for quantum
cryptography. It is the building block for quantum multi-party secure
computations and more complicated \textquotedblleft post-cold-war
era\textquotedblright\ multi-party cryptographic protocols \cite{qi75,qi139}%
. The first QBC protocol was proposed along with the very first proposal for
quantum key distribution (QKD), i.e., the Bennett-Brassard (BB) 84 protocol
\cite{qi365}. But it was pointed out at the same time that the protocol is
insecure against coherent attacks. An improved one was proposed later, known
as the Brassard-Cr\'{e}peau-Jozsa-Langlois (BCJL) 93 protocol \cite{qi43}.
It was accepted as secure for a while until a cheating strategy was found in
1996 \cite{qi74}. Shortly after, it was further concluded that any QBC
protocol cannot be unconditionally secure in principle \cite%
{*qbc27,qi24,qi23}. This result was called the Mayers-Lo-Chau (MLC) no-go
theorem. It was considered as putting a serious drawback on quantum
cryptography. Though the result is widely accepted nowadays, there is also
doubt on the generality of the theoretical model of QBC used in the no-go
proof, as it seems unconvincing that limited mathematical formulation can
characterize all possible protocols \cite{qi625}. New protocols attempting
to evade the no-go theorem were proposed every now and then \cite{qi94}-\cite%
{qi781}, though most of them turned out to be unsuccessful \cite%
{Srikanth,Yuen} or at least failed to gain a wide recognition. Nevertheless,
these attempts stimulated the research on proving the no-go theorem in more
rigorous forms. Refs. \cite{qi56,qi105,qi58,*qbc28,qi611} reviewed the
original no-go proof with fuller explanations, with some simple examples of
insecure protocols given in \cite{qi56,qi611}. Ref. \cite{qi58} also
extended the proof to cover ideal quantum coin tossing. More complicated
examples on how to apply the no-go proof to break some quantum as well as
classical bit commitment (BC) protocols which looked promising at that time
were provided in \cite{qi82} and \cite{qi47}, respectively. Refs. \cite%
{qi147,qbc49,qbc31} further studied the security bounds of QBC
quantitatively, with \cite{qbc49} focused on the protocol in \cite{qi365}.
Refs. \cite{qi101,qbc3} worked on a similar direction, while focused
especially on the class of protocols in \cite{qi94}-\cite{qi96}. Later on, a
very detailed proof was presented both in the Heisenberg picture \cite{qi323}
and the Schr\"{o}dinger picture \cite{qi715}, with the intention to achieve
a more rigorous bound on the concealment-bindingness tradeoff that can apply
to all conceivable QBC protocols in which both classical and quantum
information are exchanged, including \cite{qi94,qi100,qi190,qi422,qi361}. It
was also shown that the no-go theorem remains valid in a world subject to
superselection rules \cite{qbc36,qi610,qbc35}, or for QBC associated with
secret parameters \cite{qi240,qi283}, or when the participants are
restricted to use Gaussian states and operations only \cite{qi714}. Recent
efforts also include \cite{qbc12,qbc32}, which proved the no-go theorem with
alternative methods.

As the no-go theorem became well-accepted, people started to discuss the
possibility of building BC under various security conditions, e.g.,
classical BC under relativistic settings \cite{qi44,qi582} or tamper-evident
seals \cite{qi759}, quantum relativistic BC \cite{qbc24,qbc51},
computationally secure QBC \cite{qbc4-22,qbc4-23,qbc21,qbc22}. There are
also QBC under experimental limitations, such as individual measurements
\cite{qi85,qi137} or limited coherent measurements \cite{qbc26}, misaligned
reference frames \cite{qi295}, limited or noisy quantum storage \cite%
{qi243,qbc41,qbc39,qi538,qi796,qi795}, unstability of particles \cite%
{qi297,qi787}, Gaussian operations with non-Gaussian states \cite{qbc43},
etc. \cite{qi160,qbc58,qbc60,qbc59}. Some even considered BC in post-quantum
theories \cite{qi203}-\cite{qi651}. Others proposed less secure QBC \cite%
{qbc42,qbc29}, variations of the definition of QBC, e.g., cheat-sensitive
QBC \cite{qi150,qbc50,qi442,qbc52}, conditionally secure QBC \cite{qi63},
etc. \cite{qi162,qbc4,qi669,qbc34}.

In this paper we still focus on the original QBC without these conditions.
Basing on an existing QKD scheme using orthogonal states \cite{qi858,qi860},
we show that it becomes possible to build a QBC protocol, to which the no-go
proofs do not apply. This protocol enables many other cryptographies, and is
readily implementable with currently available technology. We also address
the relationship between this finding and the Clifton-Bub-Halvorson (CBH)
theorem \cite{qi256} which tries to characterize quantum theory in terms of
information-theoretic constraints.

QKD provides an unconditionally secure method for two remote participants to
transmit secret information against any eavesdropper. Most existing QKD
schemes (e.g., \cite{qi365,qi10,qi9}) use nonorthogonal states as carriers
for the transmitted information. Since quantum mechanics guarantees that
nonorthogonal states cannot be faithfully cloned, any eavesdropping will
inevitably introduce detectable disturbance on the states. Thus the
eavesdropper will be caught once he gains a non-trivial amount of
information. For this reason, it was once believed that nonorthogonal states
are necessary for secure QKD. But Goldenberg and Vaidman managed to present
a scheme based on orthogonal states \cite{qi858}. This brilliant idea opens
yet another path for adopting more bizarre properties of quantum mechanics
for cryptography. We will use it as the base of our current work.

Generally, in both QKD and QBC the two participants are called Alice and
Bob. But in our current proposal of QBC, the actions of Bob is more similar
to that of the eavesdropper rather than the Bob in QKD. To avoid confusion,
in this paper we use the names in the following way. In QKD, the sender of
the secret information is called Alice, the receiver is renamed as Charlie
instead of Bob, and the external eavesdropper is called Eve. In QBC, the
sender of the commitment is Alice, the receiver is Bob, and there is no Eve
since QBC merely deals with the cheating from internal dishonest
participants, instead of external eavesdropping.

\section{QKD scheme based on orthogonal states}

The QKD scheme proposed in \cite{qi858} is outlined below. Consider the
ideal case where no transmission error occurs in the communication channels.
Alice encodes the bit values $0$ and $1$ she wants to transmit to Charlie,
respectively, using two orthogonal states%
\begin{eqnarray}
0 &\rightarrow &\left\vert \Psi _{0}\right\rangle \equiv (\left\vert
a\right\rangle +\left\vert b\right\rangle )/\sqrt{2},  \nonumber \\
1 &\rightarrow &\left\vert \Psi _{1}\right\rangle \equiv (\left\vert
a\right\rangle -\left\vert b\right\rangle )/\sqrt{2}.  \label{eqpsi}
\end{eqnarray}%
Here $\left\vert a\right\rangle $ and $\left\vert b\right\rangle $ are the
localized wave packets of the same qubit. When sending these states to
Charlie, two details are important for the security of the scheme. First, $%
\left\vert a\right\rangle $ and $\left\vert b\right\rangle $ are not sent
simultaneously, but separated by a fixed delay time $\tau $. The value of $%
\tau $\ should ensure that $\left\vert a\right\rangle $ reached Charlie's
site before $\left\vert b\right\rangle $ leaves Alice's site (for
simplicity, we do not study the case where $\tau $\ is further reduced, even
though it may not hurt the security), so that the two wave packets are never
present together in the transmission channels. Second, the sending time of
each $\left\vert \Psi _{0}\right\rangle $ and $\left\vert \Psi
_{1}\right\rangle $\ is random, and kept secret from Eve until $\left\vert
a\right\rangle $ already arrived.


\begin{figure*}[tbp]
\includegraphics{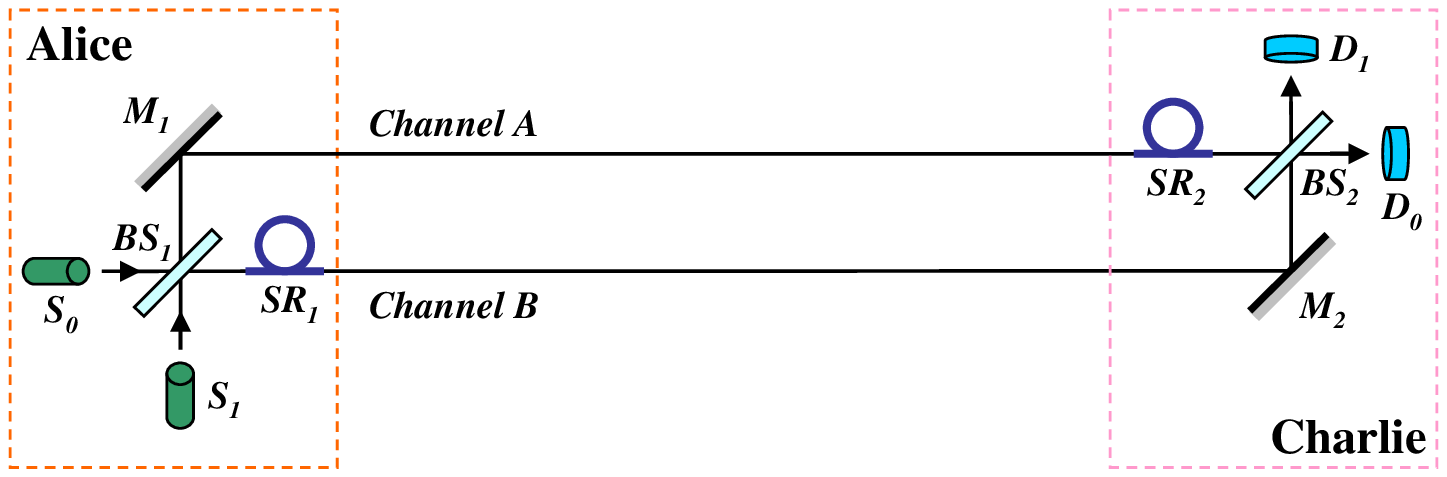}
\caption{Diagram of the experimental implementation of the QKD scheme based
on orthogonal states \protect\cite{qi858}. The state of a photon produced by
the source $S_{0}$ ($S_{1}$) will become $\left\vert \Psi _{0}\right\rangle
=(\left\vert a\right\rangle +\left\vert b\right\rangle )/\protect\sqrt{2}$\ (%
$\left\vert \Psi _{1}\right\rangle =(\left\vert a\right\rangle -\left\vert
b\right\rangle )/\protect\sqrt{2}$) after passing the beam splitter $BS_{1}$%
. The wave packets $\left\vert a\right\rangle $ and $\left\vert
b\right\rangle $ are sent through channels A and B respectively. When no
eavesdropper is present, the storage rings $SR_{1}$, $SR_{2}$ and the
mirrors $M_{1}$, $M_{2}$ will ensure the complete apparatus work as a
Mach-Zehnder interferometer with balanced arms, so that $\left\vert \Psi
_{0}\right\rangle $\ and $\left\vert \Psi _{1}\right\rangle $ will be
detected by the detectors $D_{0}$ and $D_{1}$, respectively. }
\label{fig:epsart}
\end{figure*}


FIG. 1 illustrated the diagram for an experimental implementation of
the scheme using Mach-Zehnder interferometer. Alice prepares
$\left\vert \Psi _{0}\right\rangle $ ($\left\vert \Psi
_{1}\right\rangle $) by sending a
single photon from the source $S_{0}$ ($S_{1}$), and then splits it into $%
\left\vert a\right\rangle $ and $\left\vert b\right\rangle $ using the beam
splitter $BS_{1}$. $\left\vert a\right\rangle $ is sent directly to Charlie
while $\left\vert b\right\rangle $ is delayed by the storage ring $SR_{1}$
before sending. At Charlie's site, $\left\vert a\right\rangle $ is delayed
by the storage ring $SR_{2}$ and then meets $\left\vert b\right\rangle $ at
the beam splitter $BS_{2}$ and interferes. The delay times caused by $SR_{1}$
and $SR_{2}$ are tuned equal. Thus the complete apparatus of Alice's and
Charlie's forms a balanced Mach-Zehnder interferometer, so that $\left\vert
\Psi _{0}\right\rangle $ ($\left\vert \Psi _{1}\right\rangle $) will always
make the detector $D_{0}$ ($D_{1}$) click when no eavesdropping occurs,
allowing Charlie to decode the transmitted bit value. Alice sends Charlie a
series of $\left\vert \Psi _{0}\right\rangle $ and $\left\vert \Psi
_{1}\right\rangle $, then announces all the sending times and some of the
encoded bits for security check. If all announced results match with
Charlie's measurement, the two parties keep the unannounced encoded bits as
the secret key. It was shown that the scheme is unconditionally secure \cite%
{qi858,qi860}, since Eve can never access to the entire states $\left\vert
\Psi _{0}\right\rangle $ and $\left\vert \Psi _{1}\right\rangle $, unless
she intercepts and delays $\left\vert a\right\rangle $. But then she needs
to send Charlie a \textquotedblleft dummy\textquotedblright\ state in
advance to escape the detection. However, without knowing Alice's sending
time beforehand, Eve can hardly send the dummy state at the proper time.
Thus eavesdropping will be revealed once Alice does not send any state while
Charlie's detectors click after time $\tau $.

\section{Our QBC protocol}

QBC is a two-party cryptography including two phases. In the \textit{commit}
phase, Alice (the sender of the commitment) decides the value of the bit $b$
($b=0$ or $1$) which she wants to commit, and sends Bob (the receiver of the
commitment) a piece of evidence, e.g., some quantum states. Later, in the
\textit{unveil} phase, Alice announces the value of $b$, and Bob checks it
with the evidence. An unconditionally secure QBC protocol needs to be both
\textit{binding} (i.e., Alice cannot change the value of $b$ after the
commit phase) and \textit{concealing} (Bob cannot know $b$ before the unveil
phase) without relying on any computational assumption.

To make use of the QKD scheme in \cite{qi858} for QBC, our starting point is
to treat Charlie's site as a part of Alice's, so that the two parties merge
into one. That is, Alice sends out a bit-string encoded with the above
orthogonal states, whose value is related with the bit she wants to commit.
Then she receives the states herself. Meanwhile, let Bob take the role of
Eve. His action shifts between two modes. In the \textit{intercept} mode, he
applies the intercept-resend attack to read parts of the string. In the
\textit{bypass} mode, he simply does nothing so that the corresponding parts
of the states return to Alice intact. Since the eavesdropping on every
single bit of the string has a non-trivial probability to escape Alice's
detection, at the end of the process some bits of the string become known to
Bob, while Alice does not know the exact position of these bits. Thus she
cannot alter the bit-string freely at a later time, making the protocol
binding. On the other hand, Bob cannot eavesdrop the whole string without
being detected. Thus the value of the committed bit can be made concealing
by putting a limit on the error rate Bob allowed to make in the protocol.

The rigorous description of our QBC protocol is as follows.

\bigskip

The \textit{commit} protocol:

(1) Bob chooses a binary linear $(n,k,d)$-code $C$ and announces it to
Alice, where $n$, $k$, $d$ and another parameter $s$ ($s\gg n>k>d$) are
agreed on by both Alice and Bob.

(2) Alice chooses a nonzero random $n$-bit string $r=(r_{1}r_{2}...r_{n})\in
\{0,1\}^{n}$ and announces it to Bob. This makes any $n$-bit codeword $%
c=(c_{1}c_{2}...c_{n})$ in $C$ sorted into either of the two subsets $%
C_{(0)}\equiv \{c\in C|c\odot r=0\}$ and $C_{(1)}\equiv \{c\in C|c\odot
r=1\} $. Here $c\odot r\equiv \bigoplus\limits_{i=1}^{n}c_{i}\wedge r_{i}$ .

(3) Now Alice decides the value of the bit $b$ that she wants to commit.
Then she chooses a codeword $c$ from $C_{(b)}$ randomly.

(4) Alice and Bob treat the timeline as a series of discrete time instants $%
t_{1}$, $t_{2}$, $...$, $t_{s}$ with equal intervals. Alice encodes each bit
of $c$ as $c_{i}\rightarrow \left\vert \Psi _{c_{i}}\right\rangle \equiv
(\left\vert a_{i}\right\rangle +(-1)^{c_{i}}\left\vert b_{i}\right\rangle )/%
\sqrt{2}$ and sends them to Bob. The time $t(i)$ for sending each $%
\left\vert \Psi _{c_{i}}\right\rangle $ is randomly chosen among $t_{1}$, $%
t_{2}$, $...$, $t_{s}$, while all $t(i)$'s ($i=1,2,...,n$) should be chosen
in the sequence of $i$, i.e., there should be $t(i_1)<t(i_2)$ for any $%
i_1<i_2$. Also, just as the QKD scheme in \cite{qi858}, the two wave packets
$\left\vert a_{i}\right\rangle $ and $\left\vert b_{i}\right\rangle $ of the
same qubit $\left\vert \Psi _{c_{i}}\right\rangle $\ are not sent
simultaneously. When we say that $\left\vert \Psi _{c_{i}}\right\rangle $ is
sent at time $t(i)$, we mean that $\left\vert a_{i}\right\rangle $ is sent
at time $t(i)$, while $\left\vert b_{i}\right\rangle $ is delayed and then
leaves Alice's site at time $t(i)+\tau $. The delay time $\tau $ is fixed
for all $\left\vert \Psi _{c_{i}}\right\rangle $'s and known to Bob.

(5) At each of the time instants $t_{1}$, $t_{2}$, $...$, $t_{s}$, Bob
chooses the intercept mode with probability $\alpha $ and the bypass mode
with probability $1-\alpha $.

If he chooses to apply the intercept mode at time $t_{j}$\ ($j\in
\{1,2,...,s\}$), he prepares a qubit in the state $\left\vert \Psi
_{0}\right\rangle =(\left\vert a_{j}\right\rangle +\left\vert
b_{j}\right\rangle )/\sqrt{2}$, sends the wave packet $\left\vert
a_{j}\right\rangle $ to Alice at time $t_{j}$, while $\left\vert
b_{j}\right\rangle $ is temporarily delayed. Meanwhile, Bob adds a delay
circuit to the quantum communication channel A (where the wave packets $%
\left\vert a_{i}\right\rangle $'s come from Alice). At time $t_{j}+\tau $,
he combines the output of this delay circuit with the quantum communication
channel B (where the wave packets $\left\vert b_{i}\right\rangle $'s come
from Alice), and measures whether Alice has sent him $\left\vert \Psi
_{0}\right\rangle $,\ $\left\vert \Psi _{1}\right\rangle $, or nothing at
all. If the result of the measurement is $\left\vert \Psi _{0}\right\rangle $%
\ ($\left\vert \Psi _{1}\right\rangle $), he leaves his delayed $\left\vert
b_{j}\right\rangle $ unchanged (he introduces a phase shift to change $%
\left\vert b_{j}\right\rangle $ into $-\left\vert b_{j}\right\rangle $) and
sends it to Alice. In this case, Bob learned the state Alice sent at time $%
t_{j}$\ while Alice cannot detect this action with certainty. But if Bob
found nothing in his measurement, he measures (or simply discards) $%
\left\vert b_{j}\right\rangle $. In this case, Alice's detectors will click
with probability $1/2$ due to the presence of $\left\vert a_{j}\right\rangle
$, revealing that Bob is running the intercept mode.

On the other hand, if Bob chooses to apply the bypass mode at time $t_{j}$,
he simply keeps channel A intact at time $t_{j}$, and channel B intact at
time $t_{j}+\tau $. Consequently, if a state was sent from Alice at time $%
t_{j} $, it will be returned to her detectors as-is at time $t_{j}+\tau $.

(6) Alice uses the same apparatus that Bob used in the intercept mode, to
measure the output of the quantum communication channels from Bob. She
counts the total number of the states she received from Bob, and denotes it
as $n^{\prime }$. By analyzing step (5) it can be shown that $n^{\prime
}\sim \alpha (s-n)/2+n$. Thus Alice can estimate the probability of Bob
choosing the intercept mode as $\alpha \sim 2(n^{\prime }-n)/(s-n)$. Alice
agrees to continue with the protocol if $\alpha <1-d/n$, which means that
the number of $c_{i}$'s known to Bob is $\alpha n<n-d$.

(7) Alice announces all the time instants $t(i)$'s at which she sent $%
\left\vert \Psi _{c_{i}}\right\rangle $'s ($i=1,2,...,n$). Bob checks that
he indeed detected some states at each $t(i)+\tau $ and no detection was
found at other times, as long as he has chosen the intercept mode at the
corresponding time instants. This completes the commit phase.

\bigskip

The \textit{unveil} protocol:

(8) Alice announces the values of $b$ and $c=(c_{1}c_{2}...c_{n})$.

(9) Bob accepts the commitment if $c\odot r=b$ and $c$ is indeed a codeword
from $C$, and every $c_{i}$\ agrees with the state $\left\vert \Psi
_{c_{i}}\right\rangle $ he received in the intercept mode.

\bigskip

The diagram for implementing this protocol using the Mach-Zehnder
interferometer is shown in FIG. 2.


\begin{figure*}[tbp]
\includegraphics{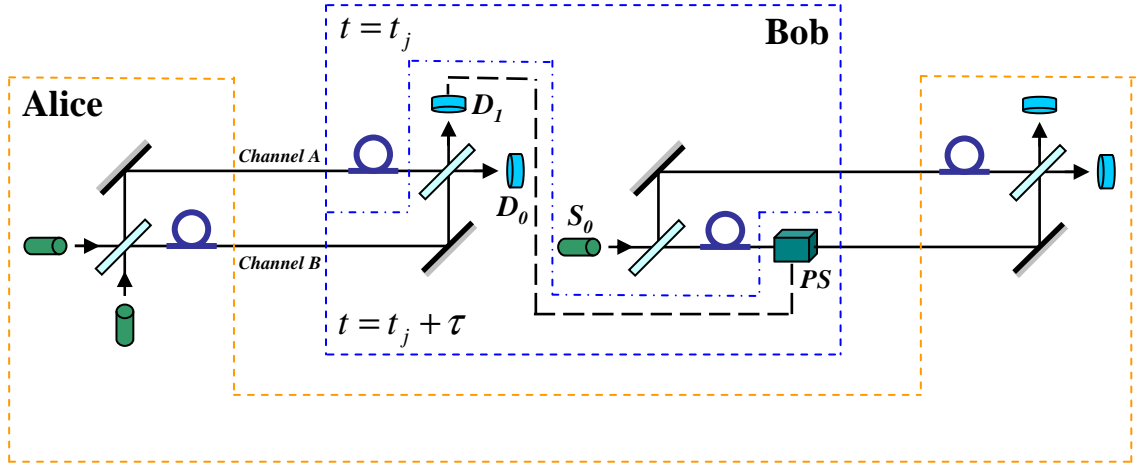}
\caption{Diagram for the apparatus of the QBC protocol when Bob chooses the
intercept mode. At time $t_{j}$, he delays anything coming from channel A,
produces $\left\vert \Psi _{0}\right\rangle =(\left\vert a\right\rangle
+\left\vert b\right\rangle )/\protect\sqrt{2}$\ using the source $S_{0}$,
and sends the wave packet $\left\vert a\right\rangle $ to Alice while
delaying $\left\vert b\right\rangle $. At time $t_{j}+\protect\tau $, he
measures the state from Alice. If the detector $D_{0}$ clicks, he sends the
delayed wave packet $\left\vert b\right\rangle $ to Alice directly. Else if
the detector $D_{1}$ clicks, he changes $\left\vert b\right\rangle $ to $%
-\left\vert b\right\rangle $ using the phase shifter PS before sending. If
none of $D_{0} $ and $D_{1}$ clicks, he discards $\left\vert b\right\rangle $%
. On the other hand, if Bob chooses the bypass mode at time $t_{j}$, he
simply removes any device in his box and let channel A (channel B) pass
through to Alice intact at time $t_{j}$ (at time $t_{j}+\protect\tau $). }
\label{fig:epsart}
\end{figure*}


Intuitively, the protocol can achieve the goal of QBC for the following
reasons. The binary linear $(n,k,d)$-code $C$ can simply be viewed as a set
of classical $n$-bit strings. Each string is called a codeword. This set of
strings has two features. (A) Among all the $2^{n}$ possible choices of $n$%
-bit strings, only a particular set of the size $\sim 2^{k}$\ is selected to
form this set. (B) The distance (i.e., the number of different bits) between
any two codewords in this set is not less than $d$. Feature (A) puts a limit
on Alice's freedom on choosing the initial state $\left\vert \Psi
_{c}\right\rangle \equiv \left\vert \Psi _{c_{1}}\right\rangle \otimes
\left\vert \Psi _{c_{2}}\right\rangle \otimes ...\otimes \left\vert \Psi
_{c_{n}}\right\rangle $. Meanwhile, feature (B) guarantees that if Alice
wants to change the string $c$ from one codeword into another, she needs to
change at least $d$ qubits of $\left\vert \Psi _{c}\right\rangle $. But the
intercept mode in the protocol enables Bob to learn about $\alpha n$\ bits
of the string $c$, while Alice does not know all the positions of these\
bits in $c$ with certainty. Therefore, when Alice alters the codeword
corresponding to $\left\vert \Psi _{c}\right\rangle $, the probability for
her to escape the detection will be only at the order of magnitude of $%
(1-\alpha )^{d}$. By increasing $d$, the security of the protocol against
Alice's cheating will be strengthened. On the other hand, feature (A) also
guarantees that the number of different codewords having less than $n-d$
bits in common increases exponentially with $k$. That is, as Bob knows only $%
\alpha n<n-d$ bits of $c$, the potential choices for $c$ are too much for
him to determine whether $c$ belongs to the subset $C_{(0)}$ or $C_{(1)}$.
Thus his knowledge on the committed bit $b$ before the unveil phase can be
made arbitrarily close to zero by increasing $k$. Fixing $k/n$ and $d/n$
while increasing $n$ will then result in a protocol secure against both
parties.

Note that when $n\rightarrow \infty $ with $k/n$, $d/n$, and $\alpha $\
fixed, the probabilities for Alice and Bob to cheat successfully in our
protocol will both drop arbitrarily close to $0$, but they never strictly
equal to $0$.
As defined in \cite{qi24,qi581}, if a protocol can make the probability of
successful cheating strictly equal to $0$, then it is considered as
\textquotedblleft perfectly secure\textquotedblright . On the other hand,
when speaking of \textquotedblleft unconditionally secure\textquotedblright
, it generally implies that the protocol should meet two requirements
simultaneously. (I) Theoretically, the security of the protocol must be
based directly on fundamental laws of physics (e.g., the validity of the
postulates of quantum mechanics or relativity) alone rather than
computational assumptions. (II) Quantitatively, the probability of
successful cheating does not equal to $0$, but can be made arbitrarily close
to $0$\ by increasing some security parameters of the protocol. To emphasize
the second meaning, some people use the term \textquotedblleft
information-theoretically secure\textquotedblright\ interchangeably with
\textquotedblleft unconditionally secure\textquotedblright\ \cite{qi610}. So
we can see that our protocol falls into this category. This is already the
best we could expect from quantum cryptography so far. For example, the
BCJL93 QBC protocol \cite{qi43} tried to reduce the probability of
successful cheating down to exactly the same level (i.e., arbitrarily close
but not equal to $0$), but proven failure by \cite{qi74}. Ref. \cite{qi147}
also showed that perfectly secure QBC is impossible. The protocols proposed
in it is even less secure, as at least one of the probabilities of Alice's
and Bob's successful cheating can never be made arbitrarily close to $0$. In
fact, even the well-known BB84 QKD protocol \cite{qi365} is not perfectly
secure. This is because the eavesdropper Eve can always perform the most
basic intercept-resend attack. That is, she intercepts any quantum state
from the sender, measure it in a basis which she chooses simply by guess,
then resends the resultant state to the receiver. While she stands a great
chance to be detected whenever her guess is wrong, we can never neglect the
probability that she can be so lucky that she guesses all the bases
correctly. Even though this probability is extremely small, and drops
arbitrarily close to $0$ with the increase of the number of states used in
the protocol, still it never strictly equal to $0$. Nevertheless, QKD is
still considered as the most secure communication method of today. Thus we
see that an unconditionally secure protocol is already good enough.

Under practical settings, some steps of our protocol may need minor
modifications. For example, the protocol can be made fault-tolerant as long
as $d/n$ is chosen to be much larger than the transmission error rate $%
\varepsilon $ of the quantum channels. This is because the distance between
any two codewords is not less than $d$. Even if a dishonest Alice replaces
the channels with noiseless ones so that she can alter up to $\varepsilon n$
bits of the string $c$ while blaming it on the transmission error, it is
still insufficient to change a codeword into another one so that her
committed bit $b$ will not be altered. For this reason, in step (9) Bob can
in fact allow the mismatched results between Alice's announced $c_{i}$ and
Bob's received $\left\vert \Psi _{c_{i}}\right\rangle $ occur with a
probability not greater than $\varepsilon $, thus makes the protocol fully
functional with noisy channels. Also, in real settings the physical systems
implementing the qubits may have other degrees of freedom, which leave rooms
for some technical cheating strategies. For instance, Alice may send photons
with certain polarization or frequency, so that she can distinguish them
from the photons Bob sends in the intercept mode. In this case, Bob and
Alice should discuss at the beginning of the protocol, to limit these
degrees of freedom to a single mode. In step (5) when Bob chooses the
intercept mode, he should also measure occasionally these degrees of freedom
of some of Alice's photons, instead of performing the measurement in the
original step (5). Then if Alice wants to send distinguishable photons with
a high probability so that they are sufficient for her cheating, she will
inevitably be detected. Another example is given in the appendix showing how
to deal with the counterfactual attack.

\section{Security}

Since the number of potential cheating strategies could be infinite, in this
work we do not attempt to prove that our protocol is unconditionally secure
against any strategy. What will be shown here is that our protocol is at
least not covered by the cheating strategy used in the MLC no-go theorem
that makes all previous QBC schemes insecure.

Briefly, the MLC no-go theorem and all its variations \cite%
{qi74,*qbc27,qi24,qi23}, \cite{qi56}-\cite{qbc32} have the following common
features.

(i) The reduced model. According to the no-go proofs, any QBC protocol can
be reduced to the following model. Alice and Bob together own a quantum
state in a given Hilbert space. Each of them performs unitary
transformations on the state in turns. All measurements are performed at the
very end.

(ii)\ The coding method. The quantum state corresponding to the committed
bit $b$ has the form%
\begin{equation}
\left\vert \psi _{b}\right\rangle =\sum\limits_{j}\lambda
_{j}^{(b)}\left\vert e_{j}^{(b)}\right\rangle _{A}\otimes \left\vert
f_{j}^{(b)}\right\rangle _{B}.  \label{coding}
\end{equation}%
Here the systems $A$ and $B$ are owned by Alice and Bob respectively, $%
\{\left\vert e_{j}^{(b)}\right\rangle _{A}\}$ is an orthogonal basis of
system $A$ while $\left\vert f_{j}^{(b)}\right\rangle _{B}$'s\ are not
necessarily orthogonal to each other.

(iii) The concealing condition. To ensure that Bob's information on the
committed bit is trivial before the unveil phase, any QBC protocol secure
against Bob should satisfy%
\begin{equation}
\rho _{0}^{B}\simeq \rho _{1}^{B},  \label{eqconcealing}
\end{equation}%
where $\rho _{b}^{B}\equiv Tr_{A}\left\vert \psi _{b}\right\rangle
\left\langle \psi _{b}\right\vert $ is the reduced density matrix of the
state sent to Bob corresponding to Alice's committed bit $b$. Note that in
some presentation of the no-go proofs (e.g. \cite{qi147,qbc31,qi323,qi714}),
this feature was expressed using the trace distance or the fidelity instead
of the reduced density matrices, while the meaning remains the same.

(iv) The cheating strategy. As long as Eq. (\ref{eqconcealing}) is
satisfied, there exists a local unitary transformation for Alice to map $%
\left\vert \psi _{0}\right\rangle $ into $\left\vert \psi _{1}\right\rangle $
successfully with a high probability \cite{qi73}. Thus a dishonest Alice can
unveil the state as either $\left\vert \psi _{0}\right\rangle $ or $%
\left\vert \psi _{1}\right\rangle $\ at her will with a high probability to
escape Bob's detection. For this reason, a concealing QBC protocol cannot be
binding.

The key that makes our protocol evade the no-go proofs is that it does not
have the feature (iii). As shown in Eq. (\ref{eqpsi}), every bit value $%
c_{i} $ in our protocol\ are encoded with orthogonal states. Therefore the
state $\left\vert \Psi _{c}\right\rangle \equiv \left\vert \Psi
_{c_{1}}\right\rangle \otimes \left\vert \Psi _{c_{2}}\right\rangle \otimes
...\otimes \left\vert \Psi _{c_{n}}\right\rangle $ corresponding to a
codeword $c$ is orthogonal to any other state $\left\vert \Psi _{c^{\prime
}}\right\rangle \equiv \left\vert \Psi _{c_{1}^{\prime }}\right\rangle
\otimes \left\vert \Psi _{c_{2}^{\prime }}\right\rangle \otimes ...\otimes
\left\vert \Psi _{c_{n}^{\prime }}\right\rangle $ corresponding to a
different codeword $c^{\prime }$. Consequently, the two Hilbert spaces
supported by the states corresponding to the codeword subsets $C_{(0)}$ and $%
C_{(1)}$ respectively are completely orthogonal to each other. Therefore it
is obvious that our protocol satisfies $\rho _{0}^{B}\perp \rho _{1}^{B}$\
instead of Eq. (\ref{eqconcealing}). Then Alice's cheating strategy (iv)
will no longer apply because the corresponding unitary transformation does
not exist without Eq. (\ref{eqconcealing}). Since all existing no-go proofs
of unconditionally secure QBC \cite{qi74,*qbc27,qi24,qi23}, \cite{qi56}-\cite%
{qbc32} have the feature $\rho _{0}^{B}\simeq \rho _{1}^{B}$, we can see
that they all fail to cover our protocol.

Let us elaborate in more details. The existence of Alice's cheating strategy
in the no-go proofs is backed by the Hughston-Jozsa-Wootters (HJW) theorem
\cite{qi73} basing on Schmidt decomposition. Following the manner of \cite%
{qi56}, it can be expressed in simple words as:

\textit{The HJW theorem: Let }$f_{1}$\textit{, }$f_{2}$\textit{, ..., }$%
f_{m} $\textit{\ and }$f_{1}^{\prime }$\textit{, }$f_{2}^{\prime }$\textit{,
..., }$f_{n}^{\prime }$\textit{\ be two sets of possible quantum states with
associated probabilities described by an identical density matrix }$\rho $%
\textit{. It is possible to construct a composite system }$A\otimes B$%
\textit{\ such that }$B$\textit{\ alone has density matrix }$\rho $\textit{\
and such that there exists a pair of measurements }$M$\textit{, }$M^{\prime
} $\textit{\ with the property that applying }$M$\textit{\ (resp. }$%
M^{\prime } $\textit{) to }$A$\textit{\ yields an index }$j$\textit{\ of
state }$f_{j}$\textit{\ (resp. }$f_{j}^{\prime }$\textit{) to which }$B$%
\textit{\ will have collapsed.}

Now consider a QBC protocol which requires Alice to encode the committed $b$
in the state%
\begin{equation}
\left\vert \psi _{b=0}\right\rangle =\sum\limits_{j}\lambda
_{j}^{(0)}\left\vert e_{j}^{(0)}\right\rangle _{A}\otimes \left\vert
f_{j}^{(0)}\right\rangle _{B},  \label{b0}
\end{equation}%
or
\begin{equation}
\left\vert \psi _{b=1}\right\rangle =\sum\limits_{j}\lambda
_{j}^{(1)}\left\vert e_{j}^{(1)}\right\rangle _{A}\otimes \left\vert
f_{j}^{(1)}\right\rangle _{B},  \label{b1}
\end{equation}%
respectively, where the meaning of the notations are the same as that of Eq.
(\ref{coding}). When the concealing condition $\rho _{0}^{B}\simeq \rho
_{1}^{B}$ is satisfied, according to the HJW theorem there exists another
basis $\{\left\vert e_{j}^{\prime }\right\rangle _{A}\}$ of system $A$ with
which we can rewrite Eq. (\ref{b0}) as
\begin{equation}
\left\vert \psi _{b=0}\right\rangle =\sum\limits_{j}\lambda _{j}^{\prime
}\left\vert e_{j}^{\prime }\right\rangle _{A}\otimes \left\vert
f_{j}^{(1)}\right\rangle _{B}.  \label{b01}
\end{equation}%
Comparing with Eq. (\ref{b1}), we can see that $\left\vert \psi
_{b=0}\right\rangle $ differs from $\left\vert \psi _{b=1}\right\rangle $\
only by a local unitary transformation $U_{A}$ of Alice which maps $%
\{\left\vert e_{j}^{\prime }\right\rangle _{A}\}$ into $\{\left\vert
e_{j}^{(1)}\right\rangle _{A}\}$. That is, with this transformation, Alice
can alter the commitment in the unveil phase by herself. The actual cheating
procedure is as follows. Alice always uses $\left\vert \psi
_{b=0}\right\rangle $ to execute the commit protocol regardless the value of
$b$. Later, if she wants to unveil $b=0$, she simply measures system $A$ in
the basis $\{\left\vert e_{j}^{(0)}\right\rangle _{A}\}$\ to collapse system
$B$ into a\ certain $\left\vert f_{j}^{(0)}\right\rangle _{B}$\ (where $j$
is determined by the quantum uncertainty in the measurement). Else if she
wants to unveil $b=1$, she rotates her basis to $\{\left\vert e_{j}^{\prime
}\right\rangle _{A}\}$ so that the corresponding measurement can collapse
system $B$ to a certain $\left\vert f_{j}^{(1)}\right\rangle _{B}$. Even if
she is required to transfer system $A$ to Bob for verification, all she
needs to do is to further apply the local unitary transformation $U_{A}$ on
system $A$ to rotate $\left\vert e_{j}^{\prime }\right\rangle _{A}$\ into $%
\left\vert e_{j}^{(1)}\right\rangle _{A}$. Thus she can always unveil $b=0$
successfully with the probability $100\%$, while unveiling $b=1$ can also be
successful with a very high probability (which can reach $100\%$ when $\rho
_{0}^{B}$ equals to $\rho _{1}^{B}$ exactly). Namely, Alice can cheat
because there are two different bases for system $A$, both of which can lead
to a legitimate outcome in the unveil phase.

But in our QBC protocol, as the state $\left\vert \Psi _{c}\right\rangle $
sent to Bob satisfies $\rho _{0}^{B}\perp \rho _{1}^{B}$, $\left\vert \psi
_{b=0}\right\rangle $ can no longer be expressed as the superposition of the
components of $\left\vert \psi _{b=1}\right\rangle $ like Eq. (\ref{b01}).
Consequently, even if Alice introduces an ancillary system $A$ entangled
with many different $\left\vert \Psi _{c}\right\rangle $'s in the form of
Eq. (\ref{coding}), there will be no alternative basis for Alice to alter
her commitment. Instead, unveiling $b=0$ and $b=1$ will be performed in the
same basis. This can be seen from the following analysis. Let $H$ denote the
Hilbert space of the composite system $A\otimes B$\ supported by all
possible committed states. Let $H_{0}$ ($H_{1}$) be its subspace supported
by all the states encoding $b=0$ ($b=1$), with $\{\left\vert
g_{j}^{(0)}\right\rangle _{A\otimes B}\}$\ ($\{\left\vert
g_{j}^{(1)}\right\rangle _{A\otimes B}\}$) denoting one of its basis. The
condition $\rho _{0}^{B}\perp \rho _{1}^{B}$ indicates that $H_{0}$ and $%
H_{1}$ have no overlap at all. Therefore $\{\left\vert
g_{j}^{(0)}\right\rangle _{A\otimes B}\}$\ and $\{\left\vert
g_{j}^{(1)}\right\rangle _{A\otimes B}\}$ share no state in common. Any
Alice's local unitary transformation $U_{A}$ on system $A$\ can be extended
as $U\equiv U_{A}\otimes I_{B}$, which becomes a unitary transformation on
the composite system $A\otimes B$. Here $I_{B}$\ is the identity operator on
system $B$. Obviously any $U$ in this form cannot map $\{\left\vert
g_{j}^{(0)}\right\rangle _{A\otimes B}\}$\ into $\{\left\vert
g_{j}^{(1)}\right\rangle _{A\otimes B}\}$. Thus Alice's actions for
unveiling $b=0$ and $b=1$, respectively, are not related with each other by
a local unitary transformation of her own. Instead, the set $\{\left\vert
g_{j}^{(0)}\right\rangle _{A\otimes B},\left\vert g_{j}^{(1)}\right\rangle
_{A\otimes B}\}=\{\left\vert g_{j}^{(0)}\right\rangle _{A\otimes B}\}\cup
\{\left\vert g_{j}^{(1)}\right\rangle _{A\otimes B}\}$ forms a single
complete orthogonal basis of the global space $H=H_{0}\oplus H_{1}$, as
either of $\{\left\vert g_{j}^{(0)}\right\rangle _{A\otimes B}\}$ and $%
\{\left\vert g_{j}^{(1)}\right\rangle _{A\otimes B}\}$ alone is incomplete.
Therefore, when writing out the Schmidt decomposition of the committed state
in forms of Eqs. (\ref{b0}) and (\ref{b1}), the states $\left\vert
f_{j}^{(0)}\right\rangle _{B}$\ and $\left\vert f_{j}^{(1)}\right\rangle
_{B} $\ belong to the same basis, instead of two different bases
nonorthogonal to each other. As a result, comparing with the description of
the HJW theorem, now $f_{1}^{(0)}$, $f_{2}^{(0)}$, ..., $f_{m}^{(0)}$ and $%
f_{1}^{(1)}$, $f_{2}^{(1)}$, ..., $f_{n}^{(1)}$ together form a single set
of orthogonal quantum states with associated probability described by a
density matrix $\rho $. When constructing a composite system $A\otimes B$\
such that $B$\ alone has density matrix $\rho $, the \textquotedblleft
two\textquotedblright\ measurements $M$, $M^{\prime }$\ (with the property
that applying $M$\ (resp. $M^{\prime }$) to $A$\ yields an index $j$\ of
state $f_{j}^{(0)}$\ (resp. $f_{j}^{(1)}$) to which $B$\ will have
collapsed) now both become incomplete measurements on system $A$. Together
they form one single complete measurement set. $\{\left\vert
e_{j}^{(0)}\right\rangle _{A}\}$ and $\{\left\vert e_{j}^{(1)}\right\rangle
_{A}\}$\ in Eqs. (\ref{b0}) and (\ref{b1}) now both belong to the same
single orthogonal basis of system $A$ corresponding to this complete
measurement. No matter what value Alice wants to unveil, her action is
always to perform the measurement in this basis. Which one of the unveiled
values will finally be obtained is determined by the form of the state Alice
prepared in the commit phase, and the quantum uncertainty in the unveil
measurement (if Alice has prepared the state in the form of Eq. (\ref%
{eqcheating}), which we will discuss in more details in the next section).
Either way, it is not determined by Alice's different actions in the unveil
phase, as there does not exist a second legitimate action at all. If Alice
insists to measure in a different basis other than $\{\left\vert
e_{j}^{(0)}\right\rangle _{A}\}\cup \{\left\vert e_{j}^{(1)}\right\rangle
_{A}\}$, it will not lead to any specific legitimate unveiled outcome with
certainty, because it will collapse the state of each qubit she sent to Bob
into $\left\vert \Psi \right\rangle =\cos \theta \left\vert a\right\rangle
+\sin \theta \left\vert b\right\rangle $\ (where $\theta \neq k\pi \pm \pi
/4 $, $k$\ is an integer) or similar forms, instead of Eq. (\ref{eqpsi}).
Thus it will only increase the probability for her cheating to be detected.
Therefore we see that the feature $\rho _{0}^{B}\perp \rho _{1}^{B}$
eliminates the existence of a second legitimate measurement basis, making
Alice's cheating strategy described in the previous paragraph futile in our
protocol.

In fact, similar characters can also be found in a bit commitment protocol
proposed by Kent \cite{qi44}, which bases its security on relativity instead
of quantum mechanics. As pointed out in the 3rd paragraph of the
introduction of \cite{qbc35}, \textquotedblleft Kent's relativistic bit
commitment protocol does not rely on the existence of alternative
decompositions of a density operator, and so its security is not challenged
by the Mayers-Lo-Chau result.\textquotedblright\ As our protocol uses
orthogonal states to encode the committed bit, it does not rely on
alternative decompositions either. Thus it can evade the MLC theorem for the
same reason.

On the other hand, our protocol is still concealing against Bob despite that
$\rho _{0}^{B}\perp \rho _{1}^{B}$. The MLC theorem suggests that protocols
satisfying this condition cannot be secure, because Bob can always perform a
measurement which optimally distinguishes $\rho _{0}^{B}$ and $\rho _{1}^{B}$%
, thus learns the value of $b$ without Alice's help. But in our protocol,
even though $\rho _{0}^{B}$ and $\rho _{1}^{B}$ are distinguishable
theoretically as the states are orthogonal, Bob is unable to perform the
corresponding measurement before the unveil phase while escaping Alice's
detection. This is because
the protocol puts a limit on the number of qubits that he is allowed to
measure, as he is required to apply the intercept mode with probability $%
\alpha <1-d/n$ only. So the key question is whether a dishonest Bob can make
his intercept mode indistinguishable with the bypass mode to Alice with a
probability higher than it was evaluated in step (5) of our protocol. This
is prevented by two important features of the QKD scheme \cite{qi858} on
which our QBC protocol is based. First, the use of the storage rings makes
the two wave packets of each single qubit of Alice never presented
simultaneously in the quantum channels. This prevents Bob from knowing the
arrival of Alice's qubit in time by measuring channel A alone, as it will
disturb the state of the qubit and make the intercept mode lose its
advantage of distinguishing Alice's state. Secondly, Alice's sending time is
random and kept secret until step (7). Therefore in step (5) Bob has to
decide himself whether to send $\left\vert \Psi _{0}\right\rangle $\ into
the quantum channels to Alice, before he can be sure whether he will detect
a qubit in the quantum channels from Alice. He cannot avoid the case where
he sent $\left\vert \Psi _{0}\right\rangle $\ to Alice, while finds out
later that Alice has not sent him a qubit at the corresponding time instant.
Then his interception will be revealed once Alice detects $\left\vert \Psi
_{0}\right\rangle $, just as expected in the protocol. Thus a dishonest Bob
intercepting more qubits than allowed will inevitably introduce a very high
estimated value of $\alpha $ in step (6), so that the cheating will be
revealed.

More generally, if there exists a strategy enabling Bob to intercept most of
Alice's qubit without being detected, then in the QKD scheme \cite{qi858},
an eavesdropper will be able to apply the same strategy to gain a
non-trivial amount of information of the secret key while escaping the
detection too. But there were already many studies on the scheme in \cite%
{qi858} proving that it is indeed unconditionally secure \cite%
{qi858,qi860,qi859,qi889,qi917}. Therefore all these proofs can be regarded
as further supports on the security of our protocol.

In short, the use of orthogonal states make our protocol evade Alice's
cheating strategy suggested by the no-go proofs, while the security against
Bob is provided by the security of the QKD scheme on which our QBC protocol
is based.

\section{Limitations and applications}

Nevertheless, our protocol has the limitation that it cannot force Alice to
commit to a classical bit. Alice can skip step (3). Then in step (4),
instead of choosing a particular codeword $c$ and preparing the system $B$
to be sent to Bob in the state $\left\vert \Psi _{c}\right\rangle \equiv
\left\vert \Psi _{c_{1}}\right\rangle \otimes \left\vert \Psi
_{c_{2}}\right\rangle \otimes ...\otimes \left\vert \Psi
_{c_{n}}\right\rangle $, she introduces an ancillary system $A$ and prepares
the state of the incremented system $A\otimes B $ in an entangled form as

\begin{eqnarray}
\left\vert A\otimes B \right\rangle &=&\sum\limits_{c\in C}\lambda
_{c}\left\vert e_{c}\right\rangle \otimes \left\vert \Psi _{c}\right\rangle
\nonumber \\
&=&\sum\limits_{c\in C_{(0)}}\lambda _{c}\left\vert e_{c}\right\rangle
\otimes \left\vert \Psi _{c}\right\rangle +\sum\limits_{c\in C_{(1)}}\lambda
_{c}\left\vert e_{c}\right\rangle \otimes \left\vert \Psi _{c}\right\rangle .
\nonumber \\
&&  \label{eqcheating}
\end{eqnarray}%
Here $\{\left\vert e_{c}\right\rangle \}$ is a set of orthogonal states that
forms a basis of system $A$. Alice keeps system $A$ at her side unmeasured,
and sends system $B$ to Bob to complete the rest of the commit protocol. By
the time she needs to unveil the committed $b$, she completes the
measurement on system $A$ and knows which $\left\vert \Psi _{c}\right\rangle
$\ system $B$ collapsed to. With this method, she can learn what can be
announced as the value of the codeword $c$ (and therefore $b$) without
conflicting with Bob's measurement. As a consequence, her commitment was
kept at the quantum level until the unveil phase. But we must note that this
problem, according to Sec. III of \cite{qi581}, \textquotedblleft is not
considered a security failure of a quantum BC protocol \textit{per se}%
\textquotedblright . This is because, as we shown above, our protocol has
the feature $\rho _{0}^{B}\perp \rho _{1}^{B}$, i.e., all $\left\vert \Psi
_{c}\right\rangle $'s corresponding to the codewords $c\in C_{(0)}$\ are
orthogonal to these corresponding to $c\in C_{(1)}$. Thus the probability
for the state Eq. (\ref{eqcheating}) to be unveiled as $b=0$ successfully is%
\begin{equation}
p_{0}=\sum\limits_{c\in C_{(0)}}\left\vert \lambda _{c}\right\vert ^{2},
\end{equation}%
while the probability for it to be unveiled as $b=1$ is%
\begin{equation}
p_{1}=\sum\limits_{c\in C_{(1)}}\left\vert \lambda _{c}\right\vert ^{2}.
\end{equation}%
The normalization condition for Eq. (\ref{eqcheating}) gives%
\begin{equation}
p_{0}+p_{1}=1.  \label{eqnormalization}
\end{equation}%
Therefore, despite that our protocol cannot force Alice to commit to a
particular classical value of $b$, she is forced to commit to a probability
distribution $(p_{0},p_{1})$ once she prepared the state of $A\otimes B $\
in step (4). She can no longer change the value of either $p_{0}$ or $p_{1} $
later. The final value of the unveiled $b$ is completely out of her control.
Instead, it is determined by the quantum uncertainty in her final
measurement on the system $A$. As stated clearly in \cite{qi581}, when Eq. (%
\ref{eqnormalization}) is satisfied, the protocol already meets the
requirement of what is defined as unconditionally secure QBC. Note that the
relativistic bit commitment protocol \cite{qi44} is well-accepted as being
unconditionally secure, even though it has exactly the same problem. Most
previous QBC protocols are considered insecure because the corresponding $%
p_{0}+p_{1}$ is larger and cannot be made arbitrarily close to $1$. In fact
in some of these protocols (e.g., \cite{qi365,qi43}), $p_{0}+p_{1}$ even
reaches or is arbitrarily close to $2$. On the other hand, if a protocol can
force Alice to commit to a particular classical $b$, i.e., besides $%
p_{0}+p_{1}=1$, both $p_{0}$ and $p_{1}$ can only take the values $0$ or $1$
instead of any value in between, then it is called a bit commitment with a
certificate of classicality (BCCC) \cite{qi581}. Namely, our protocol is a
QBC but not a BCCC.

The difference between QBC and BCCC makes it important to re-examine the
relationship between BC and other cryptographic tasks at the quantum level.
For example, though BC and oblivious transfer (OT) \cite{qi139,qbc9} are
equivalent at the classical level, our QBC protocol may not lead to
unconditionally secure quantum OT (QOT) \cite{qi75,qi52}, at least, not in
the traditional way described in these references. Note that there are many
variations of OT \cite{qi139}, e.g., 1-out-of-2 OT \cite{qi52,qi201}. Here
we use the original one \cite{qi75,qbc9} (also called all-or-nothing OT) as
an example. It is defined as the following process. Alice wants to transfer
a secret bit $b\in \{0,1\}$ to Bob. At the end of the protocol, either Bob
could learn the value of $b$ with the reliability (which means the
probability for Bob's output $b$ to be equal to Alice's input) $100\%$, or
he has zero knowledge on $b$. Each case should occur with the probability $%
1/2$, and which one finally happens is out of their control. Meanwhile,
Alice should learn nothing about which case takes place. According to Sec. 2
of \cite{qi75}, QOT can be built upon BC as follows.

\bigskip

The \textit{QOT} protocol:

(I) Let $\left\vert 0,0\right\rangle $\ and $\left\vert 0,1\right\rangle $\
be two orthogonal states of a qubit, and define $\left\vert 1,0\right\rangle
\equiv (\left\vert 0,0\right\rangle +\left\vert 0,1\right\rangle )/\sqrt{2}$%
, $\left\vert 1,1\right\rangle \equiv (\left\vert 0,0\right\rangle
-\left\vert 0,1\right\rangle )/\sqrt{2}$. That is, the state of a qubit is
denoted as $\left\vert a_{i},g_{i}\right\rangle $, where $a_{i}$\ represents
the basis and $g_{i}$\ distinguishes the two states in the same basis. For $%
i=1,...,n$, Alice randomly picks $a_{i},g_{i}\in \{0,1\}$\ and sends Bob a
qubit $\phi _{i}$ in the state $\left\vert a_{i},g_{i}\right\rangle $.

(II) For $i=1,...,n$, Bob randomly picks a basis $b_{i}\in \{0,1\}$ to
measure $\phi_{i}$ and records the result as $\left\vert
b_{i},h_{i}\right\rangle $. Then he commits $(b_{i},h_{i})$ to Alice using
the BC protocol.

(III) Alice randomly picks a subset $R\subseteq \{1,...,n\}$\ and tests
Bob's commitment at positions in $R$. If any $i\in R$ reveals $a_{i}=b_{i}$
and $g_{i}\neq h_{i}$, then Alice stops the protocol; otherwise, the test
result is \textit{accepted}.

(IV) Alice announces the bases $a_{i}$\ ($i=1,...,n$). Let $T_{0}$ be the
set of all $1\leq i\leq n$ with $a_{i}=b_{i}$, and $T_{1}$ be the set of all
$1\leq i\leq n$ with $a_{i}\neq b_{i}$. Bob chooses $I_{0}\subseteq T_{0}-R$%
, $I_{1}\subseteq T_{1}-R$ with $\left\vert I_{0}\right\vert =\left\vert
I_{1}\right\vert =0.24n$, and sends $\{I_{0},I_{1}\}$ in random order to
Alice.

(V) Alice picks a random $s\in \{0,1\}$, and sends $s$, $\beta
_{s}=b\bigoplus\limits_{i\in I_{s}}g_{i}$ to Bob. Bob computes $b=\beta
_{s}\bigoplus\limits_{i\in I_{s}}h_{i}$ if $I_{s}\subseteq T_{0}$; otherwise
does nothing.

\bigskip

If QBC instead of BCCC is used as the BC protocol in step (II), Bob can make
use of its limitation to enable a so-called honest-but-curious attack \cite%
{qi499,qi677,qi725,qi797}, as shown below. For each $\phi _{i}$ ($i=1,...,n$%
), Bob does not pick a classical $b_{i}$ and measure it in step (II).
Instead, he introduces two ancillary qubit systems $B_{i}$ and $H_{i}$ as
the storages for the bits $b_{i}$ and $h_{i}$, and prepares their initial
states as $\left\vert B_{i}\right\rangle =(\left\vert 0\right\rangle
_{B}+\left\vert 1\right\rangle _{B})/\sqrt{2}$ and $\left\vert
H_{i}\right\rangle =\left\vert 0\right\rangle _{H}$\ respectively. Here $%
\left\vert 0\right\rangle $ and $\left\vert 1\right\rangle $ are orthogonal.
Then he applies the unitary transformation%
\begin{eqnarray}
U_{1} &\equiv &\left\vert 0\right\rangle _{B}\left\langle 0\right\vert
\otimes \left\vert 0,0\right\rangle _{\phi }\left\langle 0,0\right\vert
\otimes I_{H}  \nonumber \\
&&+\left\vert 0\right\rangle _{B}\left\langle 0\right\vert \otimes
\left\vert 0,1\right\rangle _{\phi }\left\langle 0,1\right\vert \otimes
\sigma _{H}^{(x)}  \nonumber \\
&&+\left\vert 1\right\rangle _{B}\left\langle 1\right\vert \otimes
\left\vert 1,0\right\rangle _{\phi }\left\langle 1,0\right\vert \otimes I_{H}
\nonumber \\
&&+\left\vert 1\right\rangle _{B}\left\langle 1\right\vert \otimes
\left\vert 1,1\right\rangle _{\phi }\left\langle 1,1\right\vert \otimes
\sigma _{H}^{(x)}
\end{eqnarray}%
on the incremented system $B_{i}\otimes \phi _{i}\otimes H_{i}$. Here $I_{H}$%
\ and $\sigma _{H}^{(x)}$\ are the identity operator and Pauli matrix of
system $H_{i}$ that satisfy $I_{H}\left\vert 0\right\rangle _{H}=\left\vert
0\right\rangle _{H}$\ and $\sigma _{H}^{(x)}\left\vert 0\right\rangle
_{H}=\left\vert 1\right\rangle _{H}$, respectively. The effect of $U_{1}$ is
like running a quantum computer program that if $\left\vert
B_{i}\right\rangle =\left\vert 0\right\rangle _{B}$ ($\left\vert
B_{i}\right\rangle =\left\vert 1\right\rangle _{B}$) then measures qubit $%
\phi _{i}$ in the basis $b_{i}=0$\ ($b_{i}=1$), and stores the result $h_{i}$
in system $H_{i}$. It is different from a classical program with the same
function as no destructive measurement is really performed, since $U_{1}$ is
not a projective operator. Consequently, the bits $b_{i}$ and $h_{i}$ are
kept at the quantum level instead of being collapsed to classical values.

Bob then commits $(b_{i},h_{i})$ to Alice at the quantum level. This can
always be done in a QBC protocol which does not satisfy the definition of
BCCC. For example, to commit $b_{i}$ in our QBC protocol, Bob further
introduces two ancillary systems $E$ and $\Psi $\ and prepares the initial
state as

\begin{equation}
\left\vert E\otimes \Psi \right\rangle _{0}=N\sum\limits_{c\in
C_{(0)}}\left\vert e_{c}\right\rangle \otimes \left\vert \Psi
_{c}\right\rangle ,
\end{equation}%
where $N$ is the normalization constant. Let $U_{E\otimes \Psi }$ be a
unitary transformation on $E\otimes \Psi $\ which\ can map each $\left\vert
e_{c}\right\rangle \otimes \left\vert \Psi _{c}\right\rangle $\ ($c\in
C_{(0)}$) into a $\left\vert e_{c}\right\rangle \otimes \left\vert \Psi
_{c}\right\rangle $\ ($c\in C_{(1)}$), i.e., it satisfies $U_{E\otimes \Psi
}\left\vert E\otimes \Psi \right\rangle _{0}=N\sum\limits_{c\in
C_{(1)}}\left\vert e_{c}\right\rangle \otimes \left\vert \Psi
_{c}\right\rangle $. Bob applies the unitary transformation%
\begin{equation}
U_{2}\equiv \left\vert 0\right\rangle _{B}\left\langle 0\right\vert \otimes
I_{E\otimes \Psi }+\left\vert 1\right\rangle _{B}\left\langle 1\right\vert
\otimes U_{E\otimes \Psi }
\end{equation}%
on the incremented system $B_{i}\otimes E\otimes \Psi $, where $I_{E\otimes
\Psi }$\ is the identity operator of system $E\otimes \Psi $. As a result,
we can see that the final state of $B_{i}\otimes \phi _{i}\otimes
H_{i}\otimes E\otimes \Psi $\ will be very similar to Eq. (\ref{eqcheating})
if we view $B_{i}\otimes \phi _{i}\otimes H_{i}\otimes E$\ as system $A$.
Then Bob can follow the process after Eq. (\ref{eqcheating}) (note that now
Bob becomes the sender of the commitment while Alice becomes the receiver)
to complete the commitment of $b_{i} $ without collapsing it to a classical
value. He can do the same to $h_{i}$.

Back to step (III) of the QOT protocol. Whenever $(b_{i},h_{i})$ ($i\in R$)
are picked to test the commitment, Bob simply unveils them honestly. Since
these $(b_{i},h_{i})$ will no longer be useful in the remaining steps of the
protocol, it does not hurt Bob's cheating. Note that the rest $(b_{i},h_{i})$
($i\notin R$) are still kept at the quantum level. After Alice announced all
bases $a_{i}$\ ($i=1,...,n$) in step (IV), Bob introduces a single global
control qubit $S^{\prime }$ for all $i$, initialized in the state $%
\left\vert s^{\prime }\right\rangle =(\left\vert 0\right\rangle _{S^{\prime
}}+\left\vert 1\right\rangle _{S^{\prime }})/\sqrt{2}$, and yet another
ancillary system $\Gamma _{i}$ for each $i\in T_{0}\cup T_{1}-R$ initialized
in the state $\left\vert \Gamma _{i}\right\rangle =\left\vert 0\right\rangle
_{\Gamma }$. Then he applies the unitary transformation%
\begin{eqnarray}
U_{3} &\equiv &\left\vert 0\right\rangle _{S^{\prime }}\left\langle
0\right\vert \otimes \left\vert a_{i}\right\rangle _{B}\left\langle
a_{i}\right\vert \otimes I_{\Gamma }  \nonumber \\
&&+\left\vert 0\right\rangle _{S^{\prime }}\left\langle 0\right\vert \otimes
\left\vert \lnot a_{i}\right\rangle _{B}\left\langle \lnot a_{i}\right\vert
\otimes \sigma _{\Gamma }^{(x)}  \nonumber \\
&&+\left\vert 1\right\rangle _{S^{\prime }}\left\langle 1\right\vert \otimes
\left\vert a_{i}\right\rangle _{B}\left\langle a_{i}\right\vert \otimes
\sigma _{\Gamma }^{(x)}  \nonumber \\
&&+\left\vert 1\right\rangle _{S^{\prime }}\left\langle 1\right\vert \otimes
\left\vert \lnot a_{i}\right\rangle _{B}\left\langle \lnot a_{i}\right\vert
\otimes I_{\Gamma }
\end{eqnarray}%
on the incremented system $S^{\prime }\otimes B_{i}\otimes \Gamma _{i}$.
Here $I_{\Gamma }$\ and $\sigma _{\Gamma }^{(x)}$\ are the identity operator
and Pauli matrix of system $\Gamma _{i}$ that satisfies $I_{\Gamma
}\left\vert 0\right\rangle _{\Gamma }=\left\vert 0\right\rangle _{\Gamma }$\
and $\sigma _{\Gamma }^{(x)}\left\vert 0\right\rangle _{\Gamma }=\left\vert
1\right\rangle _{\Gamma }$,\ respectively. The effect of $U_{3}$ is to
compare $a_{i}$ with $b_{i}$ and store the result $(a_{i}\neq b_{i})\oplus
s^{\prime }$\ in $\Gamma _{i}$. Bob then measures all $\Gamma _{i}$ ($i\in
T_{0}\cup T_{1}-R$) in the basis $\{\left\vert 0\right\rangle _{\Gamma
},\left\vert 1\right\rangle _{\Gamma }\}$, takes $T_{0}$ ($T_{1}$) as the
set of all $1\leq i\leq n$ with $\left\vert \Gamma _{i}\right\rangle
=\left\vert 0\right\rangle _{\Gamma }$ ($\left\vert \Gamma _{i}\right\rangle
=\left\vert 1\right\rangle _{\Gamma }$) instead of how they are defined in
step (IV), and finishes the rest parts of the QOT protocol.

With this method, the division of $I_{0}$, $I_{1}$ are kept at the quantum
level. Let $I_{=}$ ($I_{\neq }$) denote the set corresponding to $%
a_{i}=b_{i} $ ($a_{i}\neq b_{i}$). We can see that $U_{3}$ makes $%
I_{0}=I_{=} $, $I_{1}=I_{\neq }$\ when $s^{\prime }=0$, while $I_{0}=I_{\neq
}$, $I_{1}=I_{=} $\ when $s^{\prime }=1$. Since $S^{\prime }$ was
initialized as $\left\vert s^{\prime }\right\rangle =(\left\vert
0\right\rangle _{S^{\prime }}+\left\vert 1\right\rangle _{S^{\prime }})/%
\sqrt{2}$, the actual result of step (IV) can be described by%
\begin{eqnarray}
&&\left\vert S^{\prime }\otimes (\bigotimes\limits_{i}B_{i}\otimes \phi
_{i}\otimes H_{i}\otimes E_{i}^{\prime })\right\rangle  \nonumber \\
&\rightarrow &\left\vert \Phi _{b}\right\rangle =(\left\vert 0\right\rangle
_{S^{\prime }}\otimes \left\vert I_{0}=I_{=}\vee I_{1}=I_{\neq }\right\rangle
\nonumber \\
&&+\left\vert 1\right\rangle _{S^{\prime }}\otimes \left\vert I_{0}=I_{\neq
}\vee I_{1}=I_{=}\right\rangle )/\sqrt{2},
\end{eqnarray}%
where $E_{i}^{\prime }$ stands for all the ancillary systems Bob introduced
in the process of committing $(b_{i},h_{i})$. Suppose that Bob announces $%
\{I_{0},I_{1}\}$ in their original order to Alice. i.e., he never announces
them in the order $\{I_{1},I_{0}\}$. After Alice announced $s$\ and $\beta
_{s}$\ in step (V), the systems under Bob's possession can be viewed as%
\begin{equation}
\left\vert \Phi _{b}\right\rangle =(\left\vert s\right\rangle _{S^{\prime
}}\otimes \left\vert I_{s}=I_{=}\right\rangle +\left\vert \lnot
s\right\rangle _{S^{\prime }}\otimes \left\vert fail\right\rangle )/\sqrt{2}.
\label{eqqot}
\end{equation}%
It means that if Bob measures system $S^{\prime }$ in the basis $%
\{\left\vert 0\right\rangle _{S^{\prime }},\left\vert 1\right\rangle
_{S^{\prime }}\}$\ and the result $\left\vert s^{\prime }\right\rangle
_{S^{\prime }}$ satisfies $s^{\prime }=s$, then he is able to measure the
rest systems and decode the secret bit $b$ unambiguously; else, if the
result satisfies $s^{\prime }\neq s$, then he knows that he fails to decode $%
b$. Now the most tricky part is, as the value of $s^{\prime }$ was kept at
the quantum level before system $S^{\prime }$ is measured, that at this
stage a dishonest Bob can choose not to measure $S^{\prime }$ in the basis $%
\{\left\vert 0\right\rangle _{S^{\prime }},\left\vert 1\right\rangle
_{S^{\prime }}\}$. Instead, by denoting $\left\vert b\right\rangle \equiv
\left\vert s\right\rangle _{S^{\prime }}\otimes \left\vert
I_{s}=I_{=}\right\rangle $, and $\left\vert ?\right\rangle \equiv \left\vert
\lnot s\right\rangle _{S^{\prime }}\otimes \left\vert fail\right\rangle $,\
Eq. (\ref{eqqot}) becomes $\left\vert \Phi _{b}\right\rangle =(\left\vert
b\right\rangle +\left\vert ?\right\rangle )/\sqrt{2}$ where $\left\vert
b=0\right\rangle \equiv (%
\begin{array}{ccc}
1 & 0 & 0%
\end{array}%
)^{T}$, $\left\vert b=1\right\rangle \equiv (%
\begin{array}{ccc}
0 & 1 & 0%
\end{array}%
)^{T}$, and $\left\vert ?\right\rangle \equiv (%
\begin{array}{ccc}
0 & 0 & 1%
\end{array}%
)^{T}$ are mutually orthogonal. Then according to Eq. (33) of \cite{qi499},
Bob can distinguish them using the positive operator-valued measure (POVM) $%
(E_{0},I-E_{0})$, where%
\begin{equation}
E_{0}=\frac{1}{6}\left[
\begin{array}{ccc}
2+\sqrt{3} & -1 & 1+\sqrt{3} \\
-1 & 2-\sqrt{3} & 1-\sqrt{3} \\
1+\sqrt{3} & 1-\sqrt{3} & 2%
\end{array}%
\right] .
\end{equation}%
This allows Bob's decoded $b$ to match Alice's actual input with reliability
$(1+\sqrt{3}/2)/2$. On the contrary, when Bob executes the QOT protocol
honestly, in $1/2$ of the cases he can decode $b$ with reliability $100\%$;
in the rest $1/2$ cases where he fails to decode $b$, he can guess the value
randomly, which results in a reliability of $50\%$. Thus the average
reliability in the honest case is $100\%/2+50\%/2=75\%<(1+\sqrt{3}/2)/2$.
Note that in the above dishonest strategy, in any case Bob can never decode $%
b$ with reliability $100\%$. Therefore it is debatable whether it can be
considered as a successful cheating, as the strategy does not even
accomplish what an honest Bob can do. That is why it is called \textit{honest%
}-but-curious behavior \cite{qi677,qi725}. The existence of this loophole
may actually come from the fact that in the literature, there is the lack of
a self-consistent definition of OT specifically made for the quantum case.
That is, the goal \textquotedblleft reaching reliability $100\%$ and $50\%$
with equal probabilities\textquotedblright\ may conflict with
\textquotedblleft reaching a maximal average reliability $75\%$ with
probability $100\%$\textquotedblright\ by nature, so that it seems
unrealistic to require a protocol to satisfy both goals simultaneously.
Therefore it is somewhat unfair to consider it as a limitation on the power
of quantum cryptography itself. Nevertheless, as this honest-but-curious
behavior provides Bob with the freedom to choose between accomplishing the
original goal of QOT and achieving a higher average reliability, it may
leave rooms for potential problems when we want to build even more
complicated cryptographic protocols upon such a QBC based QOT.

Despite of this limitation, our QBC protocol can still be used to build many
other \textquotedblleft post-cold-war era\textquotedblright\ multi-party
quantum cryptographic protocols. For example, since it makes committing a
single bit possible, then repeating the protocol many times immediately
enables quantum bit string commitment (QBSC) \cite{qi161}. Also, building
quantum strong coin tossing (QCT, a.k.a. quantum coin flipping) \cite{qi365}
with an arbitrarily small bias is straight forward. Alice and Bob first
execute our commit protocol. Then Bob announces a random bit $x$
classically. Finally, Alice unveils her committed bit $b$, and the two
parties accept $y\equiv b\oplus x$ as the coin tossing result. It is trivial
to show that even if Alice kept $b$ at the quantum level until the unveil
phase by using the state Eq. (\ref{eqcheating}), she cannot bias the final $%
y $ since she cannot change the probabilities $p_{0}$, $p_{1}$. Note that
these results suggest that all the existing no-go proofs of QBSC (e.g., \cite%
{qi371,qi197}) and QCT (e.g., \cite{qi58,qi145,qi817,qi76,qbc19}) are
incorrect. This is not surprising, because all these no-go proofs are also
based on some conditions similar to $\rho _{0}^{B}\simeq \rho _{1}^{B}$, or
even built directly on top of the no-go theorem of QBC, which are all
inapplicable to our case.

\section{Feasibility}

Our protocol is very feasible. The QKD scheme \cite{qi858} we based on was
already experimentally implemented recently \cite{qi889}. By comparing Figs.
1 and 2 it can clearly be seen that our QBC protocol can be implemented with
exactly the same devices in \cite{qi889}. Thus the QBSC and QCT protocols
built upon our QBC protocol are also straight forward with currently
available technology. Moreover, as mentioned in Sec. 3, the protocol can
easily be made fault-tolerant against noisy quantum channels. Therefore it
is extremely practical.

Comparing with the unconditionally secure BC protocols based on relativity
\cite{qi44,qi582,qbc24,qbc51}, our protocol reaches the same security level,
while the implementation is more convenient. This is because in all these
relativistic BC, both Alice and Bob must have agents to help them carrying
out the protocols. Therefore, it is in fact no longer a two-party
cryptography, as what BC should have been. Also, Alice and Bob must be
separated from their agents by a distance on the relativistic scale, i.e.,
they need to be so far apart that they cannot exchange information in time.
All these requirements obviously limit the application of their protocols.

In \cite{qi917} a variation of the QKD scheme in \cite{qi858} was
proposed, which replaced the symmetric (equal transmissivity and
reflectivity) beam splitters $BS_{1}$ and $BS_{2}$ in our FIG. 1
with asymmetric ones. The advantage is that the sending time of the
qubits no longer needs to be random. The same idea may also apply to
our protocol to bring the same advantage.

However, it is important to note that the beam splitters can be
half-silvered mirrors or similar types, but must not be polarizing
beam splitters. This is because the QKD scheme \cite{qi858} we based
on will become insecure if polarizing beam splitters are used. Let
$\left\vert H\right\rangle $ ($\left\vert V\right\rangle $) denote
the horizontally (vertically) polarized state that will always be
transmitted (reflected) by polarizing beam splitters. Eve can simply
use the same device of Charlie to measure all states come from
Alice. Then, depending on which one of her detectors clicks, she can
send $\left\vert H\right\rangle $ ($\left\vert V\right\rangle $) to
Charlie through channel B (in FIG. 1) only, let alone channel A.
This can make Charlie's detector $D_{1}$ ($D_{0}$) click with
certainty, so that Charlie always receives the same result as hers
and therefore her cheating can be covered. But if half-silvered
mirrors or similar types of beam splitters are used, when Eve sends
a state to Charlie through channel B alone, both of Charlie's
detector $D_{1}$ and $D_{0}$ will have non-vanishing probabilities
to click so that Eve cannot control the result with certainty. Then
the eavesdropping will not be successful, just as shown in the
security proof in \cite{qi858}.

\section{Relationship with the CBH theorem}

The above result is also useful for developing the understanding on
fundamental theories. The CBH theorem \cite{qi256} is an attractive attempt
to raise some information-theoretic constraints to the level of fundamental
laws of Nature, from which quantum theory can be deduced. These constraints
were suggested to be three \textquotedblleft no-go's\textquotedblright ,
which are (I) the impossibility of superluminal information transfer, (II)
the impossibility of perfectly broadcasting of an unknown state, and (III)
the impossibility of unconditionally secure BC. It was worked out in \cite%
{qi256} that these three constraints can jointly entail three definitive
physical characteristics of quantum theory, i.e., kinematic independence
(a.k.a. microcausality), noncommutative, and nonlocality. Meanwhile, to show
that these three characteristics and the above three information-theoretic
constraints are exactly equivalent, it is necessary to prove conversely that
the three characteristics can entail the three constraints. This was only
partly accomplished in \cite{qi256}. It was demonstrated that the first two
characteristics can entail constraints (I) and (II). What was left undone is
the derivation of constraint (III). Note that some people believe that the
problem was solved later by \cite{qbc35}. But in fact the no-go proof of QBC
in \cite{qbc35} was also based on the condition $\rho _{0}^{B}\simeq \rho
_{1}^{B}$, which fails to cover our protocol. Thus the derivation of
constraint (III) is still incomplete. In our understanding, this situation
is yet another evidence indicating that the MLC no-go theorem of
unconditionally secure QBC is not a necessary deduction of quantum
mechanics. In fact, the reason why the MLC theorem was included in the three
constraints, simply put, is because it can entail nonlocality. As can be
seen from features (ii) and (iv) in our above brief review of the MLC
theorem, Alice can cheat in QBC only when she has the capability to
manipulate entangled states. That is, the MLC theorem can be valid only if
the physical world allows entanglement, which is a typical example of
nonlocality. However, our QBC protocol also entail nonlocality. According to
\cite{qi859}, Eq. (\ref{eqpsi}) can be rewritten using the standard
notations of quantum optics as%
\begin{eqnarray}
\left\vert \Psi _{0}\right\rangle &=&(\left\vert 0\right\rangle \left\vert
1\right\rangle +\left\vert 1\right\rangle \left\vert 0\right\rangle )/\sqrt{2%
},  \nonumber \\
\left\vert \Psi _{1}\right\rangle &=&(\left\vert 0\right\rangle \left\vert
1\right\rangle -\left\vert 1\right\rangle \left\vert 0\right\rangle )/\sqrt{2%
},
\end{eqnarray}%
where the first and second kets refer to the two quantum communication
channels, and the $0$ and $1$ inside the kets refer to the photon number.
This indicates that Alice's transmitted states in fact contain single-photon
nonlocality. The resultant QBC can be executed only when Alice has the
capability to create such nonlocality. Otherwise, if Alice merely sends both
wave packets of a photon simultaneously into the quantum communication
channels, i.e., nonlocality is not fully utilized, then Bob can easily
intercept, clone, and resend all these orthogonal states without being
detected. That is, our result indicates that QBC can be unconditionally
secure only if there is nonlocality in the physical world. This somewhat
clarifies why most previous proposed QBC protocols (e.g., \cite{qi365,qi43})
are insecure. In these protocols, if Alice wants to commit honestly, then
sending Bob pure states unentangled with any system at Alice's side is
already sufficient. Nonlocality is not entailed when these protocols are
supposed to be executed honestly. Thus, it is not surprising that a
dishonest party who is capable of manipulating entangled states can gain
more advantages than what is allowed in these protocols. On the contrary, in
our protocol an Alice who only sends unentangled pure states will no longer
be considered as honest. Nonlocality becomes a must. Thus, we can see that
no matter the MLC theorem is correct or our QBC protocol could indeed be
unconditionally secure, nonlocality is entailed in both cases. Therefore, we
tends to believe that the (im)possibility of unconditionally secure QBC is
irrelevant to the goal of characterizing quantum theory in terms of
information-theoretic constraints. To complete the CBH theorem, we may need
to seek for another information-theoretic principle as the third constraint.

\section{Summary}

We show that if a formerly proposed QKD scheme based on orthogonal states
\cite{qi858} is secure, it can be used to build a QBC protocol which remains
concealing while the reduced density matrix $\rho _{b}^{B}$ of the state Bob
received satisfies $\rho _{0}^{B}\perp \rho _{1}^{B}$. Thus it evades the
MLC no-go theorem \cite{qi74,*qbc27,qi24,qi23}, \cite{qi56}-\cite{qbc32}
which is valid for the case $\rho _{0}^{B}\simeq \rho _{1}^{B}$\ only. The
resultant QBC protocol is not a bit commitment with a certificate of
classicality; thus, it cannot lead to unconditionally secure quantum
oblivious transfer in the traditional way. But it can lead to quantum bit
string commitment and quantum strong coin tossing. This finding suggests
that a different principle other than the MLC no-go theorem is needed for
the CBH theorem to completely characterize quantum theory in terms of
information-theoretic constraints.


The work was supported in part by the NSF of China under grant No. 10975198,
the NSF of Guangdong province under grant No. 9151027501000043, and the
Foundation of Zhongshan University Advanced Research Center.

\appendix

\section{Defeating the counterfactual attack}

Though our protocol is unconditionally secure in principle, as we mentioned
at the end of Sec. III, under practical settings minor modifications may be
needed against technical attacks.

Recently a cheating strategy against counterfactual QKD protocols \cite%
{qi801,qi1026} was proposed \cite{qi1025}. Unlike general
intercept-resend attacks in which measurements are performed on the
quantum states carrying the secret information, in this strategy the
cheater makes use of quantum counterfactual effect to detect the
working modes of the devices of other participants. Thus it was
named \textquotedblleft the counterfactual attack\textquotedblright\
\cite{qi1025}. Here we will skip how it applies to QKD protocols,
while focus only on its impact on our QBC protocol.

FIG. 3 illustrates the apparatus for the attack \cite{qi1025}. The
core is a \textquotedblleft fictitious\textquotedblright\ beam
splitter (FBS) which has the following functions.

(f1) Any photon hitting the FBS from path $c$\ will be reflected with
certainty.

(f2) When the paths $a$ and $b$ are adjusted correctly, two wave packets
coming from paths $a$ and $b$ respectively will interfere and combine
together, and enter path $c$ with certainty.

(f3) Any photon hitting the FBS from path $a$ will pass through the FBS and
enter path $d$ with certainty.


\begin{figure*}[tbp]
\includegraphics{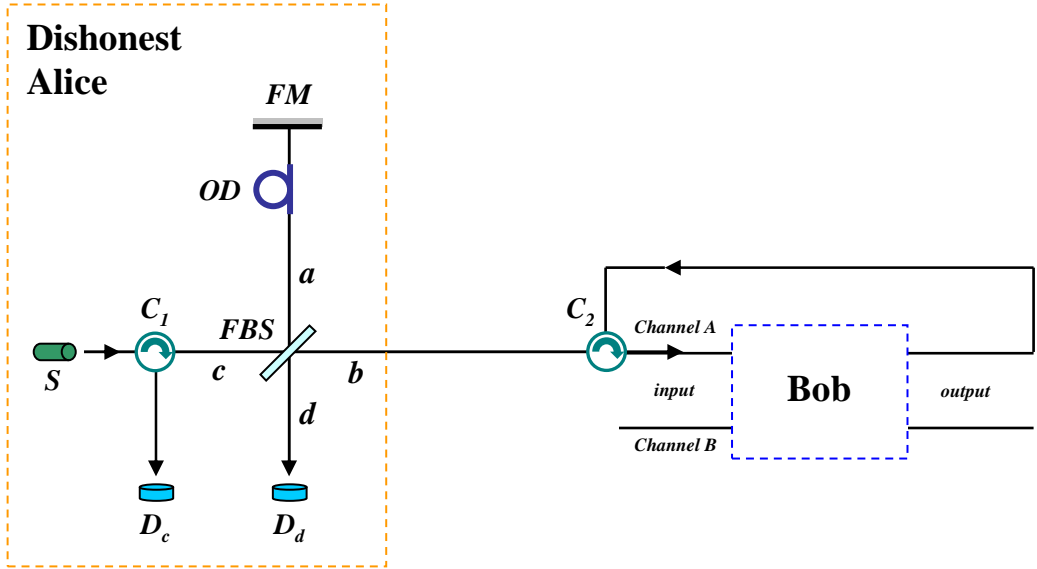}
\caption{Diagram of the apparatus for Alice's counterfactual attack.
A single-photon pulse produced by the source $S$ passes through the
optical circulator $C_{1}$ and hits the \textquotedblleft
fictitious\textquotedblright\ beam splitter (FBS) along path $c$.
Path $a$ is adjusted by the optical delay $OD$, followed by a
Faraday mirror $FM$. Any photon coming from path $c$ from the right
to the left will be detected by the detector $D_{c}$, while the
detector $D_{d}$ detects any photon coming from path $d$. Path $b$
is connected to both the input and output of Bob's channel $A$ at
time $t_{j}$\ (or both the input and output of Bob's channel $B$ at
time $t_{j}+\tau $) via the optical circulator $C_{2}$.}
\label{fig:epsart}
\end{figure*}


An ideal FBS that can realize these functions faithfully does not
exist in principle. Thus it is called \textquotedblleft
fictitious\textquotedblright . For example, devices with the
functions (f2) and (f3) may not accomplish the function (f1)
perfectly, i.e., a photon coming from path $c$ could pass the
devices with a non-trivial probability, making the attack
detectable. However, FBS can be implemented approximately by using
an infinite number of ordinary BS \cite{qi1026,qi1025}. In practice,
the number of BS involved in the implementation has to be finite.
But if the deviation from an ideal FBS is too small to be detected
within the capability of available technology, then the attack could
become a real threat.

Suppose that an ideal FBS is available to a dishonest Alice in our
QBC protocol. At each time instant $t_{j}$ (or $t_{j}+\tau $)\ in
step (4), she runs both the FBS system in FIG. 3 and the apparatus
in the honest protocol (i.e., the one shown in FIG. 2)
simultaneously in parallel, with path $b$ of the FBS system
connecting to both the input and output of Bob's channel $A$ (or
both the input and output of Bob's channel $B$). The apparatus in
FIG. 2 works as usual so that the protocol can be executed as if she
is honest, while the FBS system serves as a probe to detect Bob's
mode. According to the function (f2) of the FBS, whenever Bob
applies the bypass mode in step (5), the wave packets of a photon
Alice sent to the FBS will be returned from both paths $a$ and $b$
so that the detector $D_{c}$ will click with certainty. On the other
hand, whenever Bob applies the intercept mode, an ideal FBS can
guarantee that $D_{c}$ will never click as path $b$ is actually
blocked. Therefore Alice can learn Bob's mode unambiguously. Since
Bob does not know the state $\left\vert \Psi _{c_{i}}\right\rangle
$\ Alice sends when he applies the bypass mode, Alice can lie about
the value of the corresponding $c_{i}$ freely, thus alters her
committed $b$ in the unveil phase.

Nevertheless, it is easy to defeat this counterfactual attack. As
pointed out in Ref. \cite{qi1025}, Bob's randomizing the optical
length of path $b$ is sufficient to destroy the interference effect
in the FBS system. Therefore in our protocol, Bob can simply add
phase shifters (other than the one shown in FIG. 2) to both channels
$A$ and $B$ when he applies the bypass mode, to introduce the same
phase shift in both channels so that an honest Alice will not be
affected. Meanwhile, the amount of this phase shift is randomly
chosen and kept secret from Alice, thus she cannot know how to
adjust path $a$ to ensure $D_{c}$ clicking with certainty.
Consequently, there will be times that Alice does not know which
mode Bob is running. Then the number of $c_{i}$'s that she can alter
will be limited, which is insufficient to change the committed $b$\
as long as the value of $d/n$ in our QBC protocol is properly
chosen.


\end{document}